# An Agentic Framework for Intent Co-Creation in 6G NaaS: Architecture and Open-Source Model Evaluation

Kostis Trantzas
ECE Dept
University of Patras
GREECE
ktrantzas@ece.upatras.gr

Besiana Agko
ECE Dept
University of Patras
GREECE
besiana_agko@ac.upatras.gr

Christos Tranoris
ECE Dept
University of Patras
GREECE
tranoris@ece.upatras.gr

Irene Denazi
NetonomIQ PC
Patras
GREECE
idenazi@netonomiq.com

*Abstract*—6G network complexity necessitates high levels of autonomy, yet current intent-based systems struggle with ambiguous or incomplete human requests. This paper introduces an agent-based, intent-driven end-to-end (E2E) orchestration framework designed for Network-as-a-Service (NaaS) delivery through collaborative intent co-creation. The proposed system leverages a pool of Domain Expert Agents and a TM Forum-aligned Body-of-Knowledge (BoK) to iteratively refine user requests into deterministic, machine-readable actions. A fundamental design principle is the decoupling of cognition and actuation, where AI-driven reasoning is isolated from standardized execution controllers to ensure safety and operational trust. The framework includes a dual-layer memory system to maintain coherence during multi-step collaborations. The presented prototype, built on ETSI OpenSlice and the Model Context Protocol (MCP), evaluates across several open-source Large Language Models (LLMs). While these models demonstrate high instruction compliance, results reveal a significant gap in translating high-resolution intents into valid, catalog-backed orders without hallucinations.

*Keywords—Agentic AI; 6G networks; Intent-Based Networking; End-to-End (E2E) Orchestration; Multi-Agent Systems (MAS)*

## I. Introduction

The landscape of 6G networks is characterized by extreme heterogeneity across radio, transport, core, and cloud–edge domains. This complexity renders manual or semi-automated operations inadequate for dynamic provisioning, ultra-low latency services, and stringent reliability/sustainability targets. Consequently, network evolution must move toward high autonomy (Levels 4–5) as defined by TM Forum and 3GPP [1], [2]. Intent-Based Networking (IBN) facilitates this transition by expressing *what* outcomes are needed rather than *how* to configure the network. However, typical IBN approaches often rely on the assumption of static, well-defined intents that map directly to workflows [3]. In practice, real-world intents are frequently ambiguous or incomplete, and assurance is typically handled post-provisioning, which delays the detection of architectural mismatches and misconfigurations [4]. Agentic Artificial Intelligence (AI) offers a solution to these limitations through its ability to reason over multi-domain data, coordinate across agents, and adapt via feedback loops, but it still raises concerns around trust, governance, and alignment with industry standards [5]. To address these gaps, we propose an agentic, intent-driven end-to-end (E2E) orchestration framework centered on intent co-creation. In this framework, specialized Network as a Service (NaaS) expert agents iteratively enrich and refine user intents with operational, quality, and cost constraints, mapping them to deterministic actions only after obtaining explicit user confirmation.

## II. Related Work

Future 6G networks must orchestrate highly heterogeneous services with strict, dynamic requirements across multiple domains while still guaranteeing predictable E2E quality [6]. Despite advancements, three critical gaps remain: (i) real-world operator intents are often incomplete or ambiguous, especially for new services or runtime modifications. Assuming intents are well-formed and directly translatable leads to brittle mappings, manual interventions, or unmet expectations [7], [8]; (ii) A semantic mismatch exists between human-centric intent descriptions and machine-executable actions, as template or rule-based translators struggle to adapt to new service types, cross-domain trade-offs, and evolving conditions [9]; and (iii) quality assurance is usually treated as a separate, post-deployment activity, with Service Level Agreements (SLAs)/Service Level Objectives (SLOs) defined offline, making verification reactive and preventing quality objectives from guiding provisioning decisions upfront [10].

While Agentic AI can manage this complexity, its integration into operational orchestration remains problematic. Direct AI control lacks determinism, trust, and standards compliance required for production environments, yet restricting AI to passive analytics underutilizes its potential for reasoning, negotiation, and adaptation in E2E scenarios [11]. This motivates this research attempt to design an agentic Large Language Model (LLM)-powered framework, which interactively cocreates and validates intents, translates confirmed intents into deterministic actions, and supports continuous quality verification, all while remaining aligned with existing control systems and governance requirements.

Bridging the semantic gap between intent and action requires a deterministic reference model. Communication Service

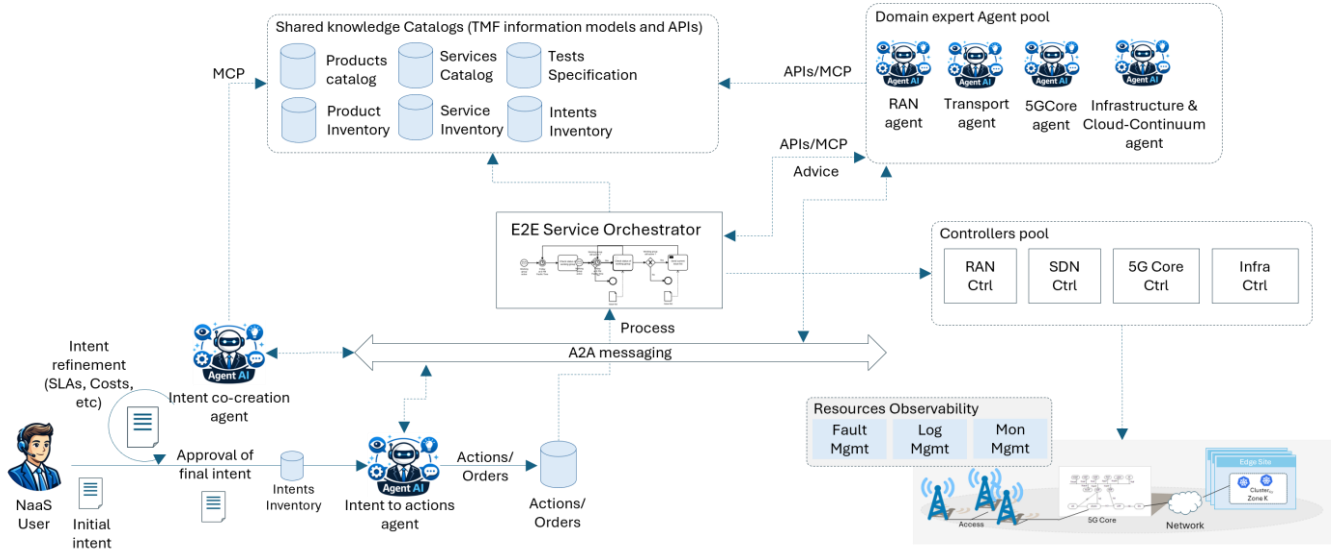

Figure 1 Proposed Agentic, Co-creation Intent-Driven E2E Orchestration

Providers (CSPs) typically commercialize NaaS as TM Forum Product Offerings (TMF620), which bundle cost, policies, and SLA terms and link to Product Specifications defining technical characteristics and parameters (e.g., location, capacity, duration) [12]. These then map to Service and Resource Specifications that realize the offer across Radio Access Network (RAN), transport, core, and edge/cloud domains, with an E2E orchestrator decomposing orders into service and resource orders and coordinating domain controllers. Product, Service, and Resource Inventories preserve runtime traceability and full lifecycle management (activation, modification, scaling, healing, termination), forming a deterministic, standards-based chain that underpins our 6G NaaS Body-of-Knowledge (BoK).

### A. Intent-Based Networking and Autonomous Networks

IBN is a key paradigm for simplifying network management by abstracting low-level configurations. Industry standardization bodies, such as TM Forum [13] and ETSI [14], along with recent research [15], position intent management frameworks as central building blocks of Autonomous Network (AN) architectures, typically using closed-loop control to translate high-level intents into concrete policies and configurations across heterogeneous domains.

Despite the significant advancements in defining intent models and interfaces, existing solutions remain largely dependent on predefined intent templates and deterministic rule-based translation engines. Consequently, they struggle to resolve incomplete or ambiguous intents, particularly in scenarios requiring a human-in-the-loop interaction. Furthermore, current models offer limited capacity for the dynamic negotiation of multi-dimensional constraints, such as trade-offs between quality-of-service (QoS), operational cost, or resource availability.

### B. AI and Agent-Based Network Management

The convergence of Agentic AI and Multi-Agent Systems (MAS) is fundamentally transforming AN control, specifically for distributed, large-scale environments, such as modern cellular networks [16]. AI techniques have been increasingly applied to network management tasks, including optimization, fault detection, and predictive analytics. More recently, agent-based and multi-agent systems have been proposed to enable distributed intelligence across network domains [17]. In these systems, autonomous agents observe the network states, reason about high-level objectives, and coordinate execution through inter-agent communication [18].

Agentic AI frameworks extend this paradigm by integrating LLMs and advanced reasoning chains, enabling agents to interpret high-level goals, negotiate with peers, and adapt their strategies over time. Research in Open RAN, AI-native networks, and AI-driven orchestration shows that such systems can significantly improve flexibility and responsiveness, but most existing work remains limited to narrow functional silos or assumes direct agent control over network functions. This unmediated control raises concerns about determinism, operational safety, and legacy compatibility, leaving open challenges in conflict mitigation, secure decision-making, and delivering explainable, trustworthy AI in production-grade orchestration [19].

### III. OUR APPROACH: AGENTIC, INTENT-DRIVEN E2E ORCHESTRATION WITH TEST-DRIVEN QUALITY ASSURANCE

### A. Design Principles

Our methodology is founded on the following principles:
- Intent Centricity: Prioritizing desired outcomes over procedural configurations.
- Agentic Autonomy: Leveraging agentic reasoning with mandatory human-in-the-loop validation to ensure operational trust and governability.
- Decoupling of Cognition and Actuation: Isolating AI-driven reasoning from the deterministic execution layer.

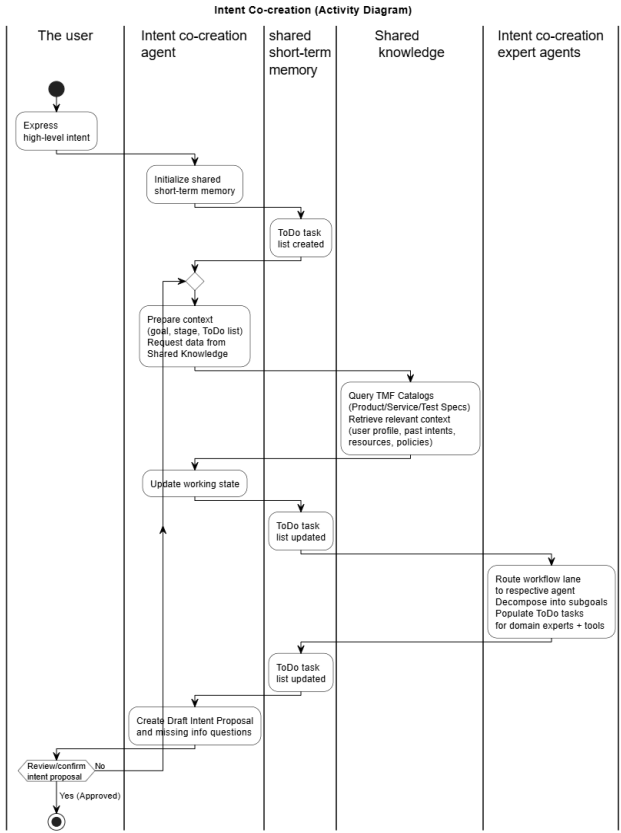

Figure 1. Agents' interaction for Co-creation of final intent

As illustrated in Fig. 1, our agentic, intent-driven E2E orchestration tightly couples intent handling with deterministic execution. A user or application intent describes desired outcomes and is treated as an evolving request rather than a static command. An *Intent Co-Creation* Agent collaborates with a pool of Domain Expert Agents over an Agent-to-Agent (A2A) message bus to iteratively refine this intent into operational information.

The Intent Co-Creation Agent acts as the primary orchestrator and gateway, as depicted in Fig. 2. This agent grounds the request against Shared Knowledge Catalogs, producing a canonical intent with explicit QoS, location, cost, and policy constraints. Once grounded, the Co-Creation Agent selects a workflow trajectory (e.g., order management, service/resource support, troubleshooting), decomposes the intent into domain-specific queries and actions, and dispatches them to Domain Expert Agents for feasibility assessments and configuration parameters. As results return, the Co-Creation Agent reconciles conflicts, records decision checkpoints in the shared case state (discussed in the upcoming Memory Management section), and produces a structured intent contract that is presented to the user for explicit confirmation.

Once confirmed, the intent contract is translated into machine-readable actions executed by the E2E orchestrator, yielding a structured plan with service decomposition, dependency ordering, and domain-specific intents. The design maintains a strict separation between cognition and actuation: agentic components perform reasoning and planning, while standardized domain controllers deterministically execute actions, ensuring AI never directly manipulates the network in an unsafe or non-deterministic way.

### B. The 6G NaaS Body-of-knowledge for creation and planning

The framework relies on a *6G NaaS BoK*, a shared machine-readable repository that acts as the system's "source of truth" and lets agents turn human intents into catalog-backed service compositions and executable plans. The BoK builds on TMF-based *Shared Knowledge Catalogs*, aggregates specifications plus site-specific constraints, and offers semantic mappings that translate natural language requests into precise performance envelopes. It also maintains dependency and policy graphs, so Co-Creation agents ground and validate intents while orchestrator and domain experts derive ordered actions and guardrails for safe execution.

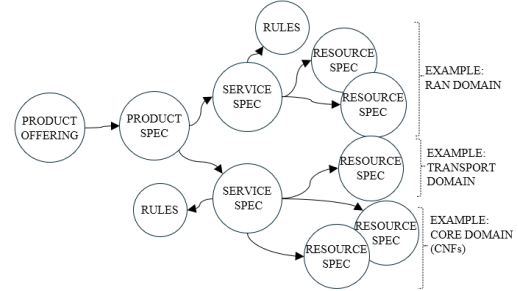

Figure 2. A typical knowledge graph from our BoK

As shown in Fig. 3, the BoK utilizes a knowledge graph to capture the hierarchical decomposition of NaaS delivery, linking commercial and technical layers. Product Offerings (what customers buy) map to Product Specifications, which in turn map to Service Specifications and finally to Resource Specifications that represent concrete capabilities in underlying domains. Policy "rules" drive these translations at each step, with resource specs grouped by domain so a single commercial product can trigger coordinated configurations across multiple network domains.

### C. Memory and state management

Our architecture maintains coherent long-term interactions by combining short-term and external long-term memory. Short-term memory is a shared, in-context *working state* (a to-do list with next actions and expected outputs, created with LLM assistance), while long-term memory is an external case file that stores the canonical catalog-backed intent, constraints, assumptions, decision log, and derived service composition.

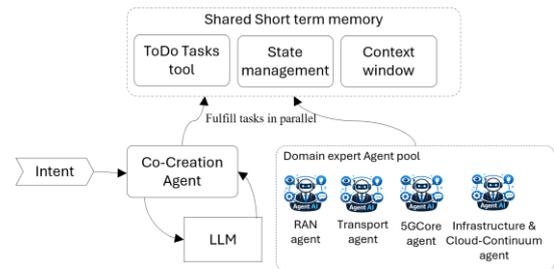

Figure 3. Shared short-term memory process for maintaining coherence

As depicted in Fig. 4, a Co-Creation Agent uses an LLM to interpret the request and decompose it into sub-tasks for Domain Expert Agents. These agents share the short-term workspace to track stage, constraints, and interim decisions; they execute tasks in parallel and feed results back to the Co-Creation Agent, which updates the shared state, reconciles constraints, and consistently drives the workflow from early decisions to final actions.

## IV. PROTOTYPE IMPLEMENTATION

### A. Extending OpenSlice for agentic-enabled orchestration

OpenSlice is an open-source, service-based Operations Support System (OSS) for delivering Network as a Service (NaaS) [20], under the operational umbrella of ETSI. It provides an end-to-end orchestration framework that allows service providers to design, order, and manage complex network services across multi-domain environments, including 5G core, radio, transport, and cloud. It aligns with ETSI, TM Forum, and 3GPP standards to bridge complex infrastructure with agile service delivery, empowering operators with plug-and-play controllers for modern telco-cloud requirements. The group operates as an Apache-licensed collaborative platform where industry and research partners prototype and validate next-generation (e.g., 6G) technologies.

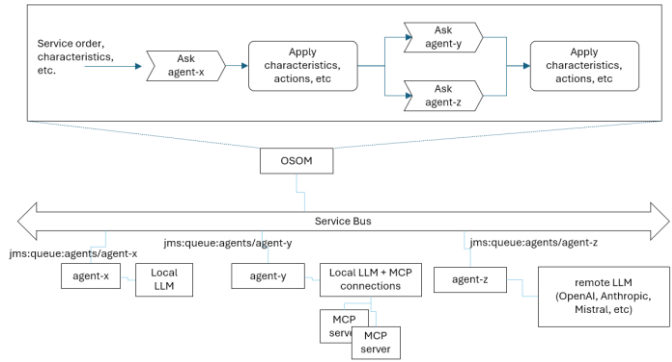

Figure 4. Message-driven, multi-agent orchestration pattern built around OpenSlice Service Orchestrator (OSOM)

### B. Prototype Agentic Design, development

#### 1) Design

Fig. 5 shows a message-driven, multi-agent orchestration pattern around OSOM (the OpenSlice E2E Service Orchestrator) and a shared Service Bus. A "service order/characteristics" request is first processed by one agent, then refined by additional agents in multiple cycles before converging into a final "apply characteristics/actions" stage, so intents and orders are progressively enriched instead of being mapped once to a static workflow. This illustrates a practical realization of our intent co-creation and iterative refinement concept.

Operationally, OSOM publishes tasks to the bus and agents subscribe asynchronously, each potentially using different inference backends (local LLMs, MCP tools, or remote providers), which enables decoupled A2A coordination, safe tool-mediated access to platform capabilities, and hybrid on-prem/remote deployments. OSOM remains the deterministic lifecycle executor, while the agents implement the cognitive layer.

#### 2) Development

We implement agents as Java microservices using the Spring Boot AI stack, consistent with existing OSL components. An example agent code is available as open source at the respective code repository [21]. Each agent runs as a message-driven service that listens on a Java Message Service (JMS) queue, treating incoming messages as prompts and enabling asynchronous invocation by OpenSlice components without synchronous Representational State Transfer (REST) chains. The LLM backend is provided via Spring AI with Ollama, via the *spring-ai-starter-model-ollama* module, supporting on-premise model execution. Architecturally, each agent combines an "endpoint", an A2A messaging backbone, and MCP-based tool access: JMS queues provide a decoupled, scalable coordination layer, while MCP clients expose controlled actuation over OpenSlice TMF Operate Application Programming Interfaces (APIs) and workflows, enforcing strict separation between cognitive reasoning and deterministic execution.

## V. EVALUATION OF PROTOTYPE AGENTS

We evaluated the prototype by measuring the decision-making accuracy of multiple open-source LLMs, each integrated via MCP with our TMF-aligned BoK, on a Kubernetes cluster with three worker nodes and an RTX A6000-equipped node for inference. This setup served as our baseline, but we observed that memory-efficient inference (e.g., AirLLM-style techniques) can significantly reduce hardware requirements and enable large models on consumer-grade edge platforms with limited impact on reasoning quality. The agents operated against the ETSI OpenSlice catalog with Product Offering that defined the technical and commercial boundaries for all evaluated decisions, as seen in Table I.

TABLE I. INDICATIVE PRODUCT OFFERINGS

| Product Offering | Tier / Type | Parameters | Unit Cost (€) |
|---|---|---|---|
| On-demand Network Slice | Platinum | cityName, sliceProfile | 1000 / day |
| | Gold | | 700 / day |
| | Silver | | 300 / day |
| Edge Media Cache Server | Large (GPU) | cityName, sliceProfile | 300 / day |
| | Large | | 200 / day |
| | Small | | 50 / day |
| Service APIs Exposure | Standard | - | 100 / day |
| Network Slice Observability | Admin Access | - | 100 / day |
| Service Setup and VPN | Standard | - | 100 (once) |

### A. Experiment setup – Agent skills & Model Selection

Before each experiment, each agent (backed by a different LLM) was given access to the BoK, tools, and a set of task-specific skills. These skills, defined in a standardized, file-based format [22], bridge the LLM's reasoning with external APIs and code, enabling the agent to recognize orchestration contexts and trigger appropriate workflows. For example, an Intent-to-Service Mapping skill takes a natural-language request (e.g., an XR stream with latency and cost constraints), extracts technical and business requirements, queries the 6G NaaS BoK, maps them to standard Service Specification characteristics, and produces a structured service plan. Once the user confirms, the agent invokes catalog, inventory, and order tools via MCP to execute the agreed plan. We constrained the co-creation agent with a strict skill policy: (i) use only catalog tools for product lookup, (ii) never map intents to products without catalog lookup, (iii) recommend only catalog products, (iv) never place orders without explicit user confirmation, (v) do not mention services. The selection of open source LLMs for this study was done based on popularity and memory constraints and governed by two primary categories: (i) reasoning-centric models with tool calling support, and (ii) tool-augmented generalist models.

### B. Performing the experiment

To evaluate the framework's ability to handle complex, multi-domain requirements, we defined a representative NaaS benchmark intent, as depicted in Fig.7.

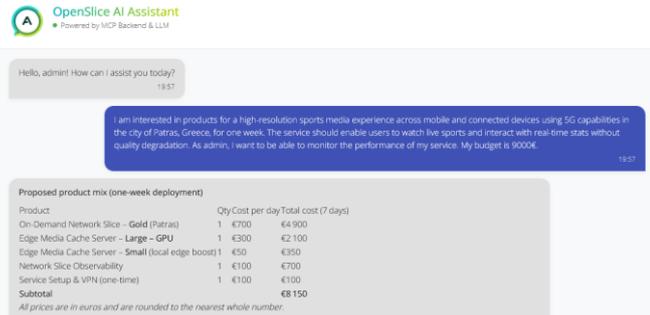

Figure 5 The OpenSlice AI Assistant Interface

The co-creation process follows a structured, multi-turn dialogue and consists of a sequential series of queries:
- Q1 (Intent Ingestion): Initial high-level intent
- Q2 (Alternative Composition): Propose alternative products that satisfy the primary constraints.
- Q3 (Product Combination Decision): The user selects a specific product combination.
- Q4 (Temporal Specification): The user defines the operational lifecycle (e.g., start date and duration).
- Q5 (Serialization and Confirmation): TMF order payload generated and awaits the user's confirmation.

To evaluate the efficacy of the agentic framework, we established a "ground truth" reference configuration, representing the optimal decisions a senior 5G engineer would make for the specified scenario. The framework's performance is measured by its ability to converge on this expert-defined baseline, across three dimensions:
1. **Functional Composition and Selection:** The agent must recover the exact four-product bundle defined by the expert baseline (On-demand Network Slice, Edge Media Cache Server, Service Setup and VPN, Slice Observability) from the catalog.
2. **Multi-dimensional Constraint Satisfaction:** The agent must not only pick the right products but also compute the one-week total cost correctly and keep the proposal within the 9,000€ budget.
3. **Deterministic Technical Initialization:** The agent must generate a valid order payload by correctly setting key parameters (e.g., geographical coverage) before handing it off to the E2E orchestrator.

TABLE II. CO-CREATION AGENT CONVERGENCE

| LLM | Correct Product Composition | Hallucinated Products | Correct Total Cost | Correct Duration | Baseline Achievement | Total Dialogue Time (min) |
|---|---|---|---|---|---|---|
| **Reasoning Models** | | | | | | |
| Gpt-oss:20b | 3 (75%) | 0 | Pass | Pass | Partial | 6 |
| Qwen3:32b | 4 (100%) | 1 | Pass | Fail | Partial | 14 |
| Qwen3-vl:8b | 3 (75%) | 0 | Fail | Fail | Fail | 26 |
| Deepseek-r1:32b | 0 (0%) | 0 | Fail | Fail | Fail | - |
| Magistral:24b | 0 (0%) | 0 | Fail | Fail | Fail | - |
| **Non-Reasoning Models** | | | | | | |
| Llama3.1:8b | 2 (50%) | 0 | Fail | Fail | Fail | 6 |
| Llama3.2:3b | 2 (50%) | 1 | Fail | Fail | Fail | 5 |
| Mistral-small3.2:24b | 3 (75%) | 1 | Pass | Fail | Partial | - |
| Ministral-3:3b | 3 (75%) | 0 | Pass | Fail | Partial | 3 |
| Granite3.1-moe:3b | 0 (0%) | 16 | Fail | Fail | Fail | - |
| Mistral:7b | 0 (0%) | 4 | Fail | Fail | Fail | - |
| Smollm2:1.7b | 0 (0%) | 4 | Fail | Fail | Fail | - |
| Mistral-nemo:12b | 0 (0%) | 0 | Fail | Fail | Fail | - |

The selected open-source LLMs were evaluated against the ground-truth configuration using six metrics: product composition correctness, number of hallucinated products, total cost accuracy, temporal reasoning, overall baseline achievement, and total dialogue time, as presented in Table II. The first four metrics quantify how well each model replicated the expert four-product bundle, avoided non-catalog items, computed the correct cost, and respected lifecycle constraints, while baseline achievement provides a qualitative rating (Pass/Partial/Fail) of reasoning and tool usage, and total dialogue time measures end-to-end latency. Overall, most models could parse the initial high-level intent, but their behavior diverged significantly during the co-creation phase (Q1–Q5), exposing substantial differences in robustness and reliability. More specifically, summarizing our findings:

- **Gpt-oss-20b:** best overall, accurate cost and date handling, no hallucinations.
- **Qwen3-32b:** moderate performance, hallucinated products, failed dates, tried to order directly.
- **Qwen3-vl-8b:** slowest model to respond, session terminated after 20 minutes without further output.
- **Deepseek-r1-32b:** toll calling not supported, despite the model's claims in Ollama.
- **Magistral-24b:** failed to provide expected products.

- **Llama3.1-8b:** hallucinated product IDs, failed dates, and cost estimations.
- **Llama3.2-3b:** prompt responses, hallucinated products, failed dates, tried to order directly.
- **Mistral-small3.2-24b:** hallucinated products, tried to order directly, which led to an unhandled error
- **Ministral-3-3b:** fastest model, no hallucinations, failed dates
- **Granite3.1-moe-3b / Mistral-7b / Smollm2:1.7b:** failed tool calling, instead hallucinated products.
- **Mistrall-nemo-12b:** failed tool calling, which led to the model not being responsive.

## VI. Discussion and Conclusions

This work shows that achieving Level 4–5 Autonomous Networks in 6G requires moving from static, template-based intent mapping to dynamic intent co-creation, where a multi-agent orchestration layer and a TM Forum-aligned Body-of-Knowledge iteratively refine ambiguous human requests into machine-ready actions. The core architectural contribution is the strict decoupling of cognition and actuation: non-deterministic LLM-based reasoning is fully isolated from standardized controllers, preserving the safety and operational trust needed in production while still exploiting LLM flexibility.

Our prototype validates intent-to-order translation and catalog adherence, leaving existing controllers to deterministically orchestrate services and resources. Experiments on ETSI OpenSlice reveal large performance differences among open-source models: Gpt-oss-20b showed zero hallucinations and full instruction compliance for NaaS delivery, whereas larger models often hallucinated products or violated budget and temporal constraints. This indicates that domain-specific grounding and fine-tuning matter more than sheer parameter count for specialized networking tasks. Future work includes optimizing the hardware footprint for constrained environments, comparing against commercial LLMs, and extending the agentic framework beyond intent-to-order translation toward full service and resource orchestration.

## Acknowledgment

This work has received funding from European Projects: COP-PILOT Grant agreement ID: 101189819, AMAZING-6G Grant agreement ID: 101192035, FIDAL Grant agreement ID: 101096146